\documentclass[aip,amsmath,amssymb,reprint,twocolumn]{revtex4-1}

\usepackage{graphicx,hyperref}

\newcommand{\uphi}{\hat{\boldsymbol{\phi}}}

\newcommand{\uZ}{\hat{\mathbf{Z}}}
\newcommand{\relphantom}[1]{\phantom{\mathrel{#1}}}
\newcommand{\abs}[1]{\left|#1\right|}
\newcommand{\jump}[1]{\left[\left[#1\right]\right]}

\begin{document}


\title{Multi-region relaxed magnetohydrodynamics with anisotropy and flow} 



\author{G.R. Dennis}
\email[]{graham.dennis@anu.edu.au}
\affiliation{Research School of Physics and Engineering, Australian National University, ACT 0200, Australia}

\author{S.R. Hudson}
\affiliation{Princeton Plasma Physics Laboratory, PO Box 451, Princeton, NJ 08543, USA}

\author{R.L. Dewar}
\author{M.J. Hole}
\affiliation{Research School of Physics and Engineering, Australian National University, ACT 0200, Australia}


\date{\today}

\begin{abstract}

We present an extension of the multi-region relaxed magnetohydrodynamics (MRxMHD) equilibrium model that includes pressure anisotropy and general plasma flows.  This anisotropic extension to our previous isotropic model is motivated by \citeauthor{Sun:1987}'s model of relaxed anisotropic magnetohydrodynamic equilibria.  We prove that as the number of plasma regions becomes infinite, our anisotropic extension of MRxMHD reduces to anisotropic ideal MHD with flow.  The continuously nested flux surface limit of our MRxMHD model is the first variational principle for anisotropic plasma equilibria with general flow fields.

\end{abstract}


\maketitle 

\section{Introduction}

The construction of magnetohydrodynamic (MHD) equilibria in three-dimensional (3D) configurations is of fundamental importance for understanding toroidal magnetically confined plasmas.  The theory and numerical construction of 3D equilibria is complicated by the fact that toroidal magnetic fields without a continuous symmetry are generally a fractal mix of islands, chaotic field lines, and magnetic flux surfaces.  \citet{Hole:2007}  have proposed a variational method for isotropic 3D MHD equilibria that embraces this structure by abandoning the assumption of continuously nested flux surfaces usually made when applying ideal MHD.  Instead, a finite number of flux surfaces are assumed to exist in a partially relaxed plasma system.  This model, termed a multi-region relaxed MHD (MRxMHD) model, is based on a generalization of the Taylor relaxation model \citep{Taylor:1974,Taylor:1986} in which the total energy (field plus plasma) is minimized subject to a finite number of magnetic flux, helicity and thermodynamic constraints.  

Obtaining 3D MHD equilibria that include islands and chaotic fields is a difficult problem, and a number of alternative approaches have been developed, including iterative approaches \citep{Reiman:1986,Suzuki:2006} and variational methods for linearized perturbations about equilibria with nested flux surfaces \citep{Hirshman:2011,Helander:2013}.  In general, variational methods have more robust convergence guarantees than iterative methods, and all else being equal, are usually preferable.  However, variational methods for plasma equilibria require constraints to be specified and enforced in order to obtain non-vacuum solutions.  The variational methods employed by \citet{Hirshman:2011} and \citet{Helander:2013} specify these constraints in terms of the flux surfaces of a nearby equilibrium with nested flux surfaces.  These methods are therefore necessarily perturbative, as opposed to the iterative methods of \citet{Reiman:1986} and \citet{Suzuki:2006} which aim to solve the full nonlinear 3D MHD equilibrium problem.  The MRxMHD model is a variational method and must also enforce constraints to obtain non-vacuum solutions.  The approach taken by MRxMHD is to assume the existence of a finite number of good flux surfaces, and to enforce plasma constraints in the regions bounded by these good flux surfaces.  This approach allows MRxMHD to solve the full nonlinear 3D MHD equilibrium problem with the assumption that there exist a finite number of flux surfaces that survive the relaxation process.  This assumption is motivated by the work of \citet{Bruno:1996}, who have proved that for sufficiently small deviations from axisymmetry such flux surfaces will exist and that they can support non-zero pressure jumps.

The MRxMHD model has seen some recent success in describing the 3D quasi-single-helicity states in RFX-mod \citep{Dennis:2013b}; however, it must be extended to include anisotropic pressure as significant anisotropy is observed in high-performance devices, particularly in the presence of neutral beam injection and ion-cyclotron resonance heating \citep{Fasoli:2007,Cooper:1980,Hole:2011a}.  Our extension of MRxMHD to include pressure anisotropy is guided by the work of \citet{Sun:1987} who studied a model for relaxed anisotropic plasmas by constraining the parallel and perpendicular entropies $S_\parallel = \int \rho \ln \left(p_\parallel B^2 / \rho^3 \right)\, d^3\tau$ and $S_\perp = \int \rho \ln \left[p_\perp / (\rho B)\right]\, d^3\tau$, in addition to the flux and magnetic helicity constraints considered by \citet{Taylor:1986}.  The model studied by \citeauthor{Sun:1987} is a special case of the single plasma-region, zero-flow limit of the anisotropic MRxMHD model presented in this paper.

In the opposite limit, as the number of plasma interfaces becomes large and the plasma contains continuously nested flux surfaces, it is desirable for anisotropic MRxMHD to reduce to anisotropic ideal MHD.  We prove this limit to be true in Sec.~\ref{sec:ContinuousLimit}, demonstrating that anisotropic MRxMHD (with flow) essentially ``interpolates'' between an anisotropic Taylor-Woltjer relaxation theory on the one hand and anisotropic ideal MHD with flow on the other.  The continuously nested flux surface limit of anisotropic MRxMHD is, to the authors' knowledge, the first variational energy principle for anisotropic plasma equilibria with general flow fields.  This is a generalization of earlier work developing variational principles for isotropic plasma equilibria with flow \citep{Hameiri:1998}.

This paper is structured as follows: in Sec.~\ref{sec:MRxMHDModel}, we give a summary of the MRxMHD model and its solution for a finite number of plasma regions before presenting our extension to include pressure anisotropy.  In Sec.~\ref{sec:ContinuousLimit}, we prove that this extension of MRxMHD reduces to anisotropic MRxMHD with flow in the limit of continuously nested flux surfaces.  This is followed by an example application of the anisotropic MRxMHD model to a reversed-field pinch (RFP) plasma in Sec.~\ref{sec:Example}.  The paper is concluded in Sec.~\ref{sec:Conclusion}.

\section{The Multi-region relaxed MHD model}
\label{sec:MRxMHDModel}

\subsection{The isotropic, zero-flow limit}
\label{sec:ZeroFlowMRxMHD}
\begin{figure}
  \includegraphics[width=8cm]{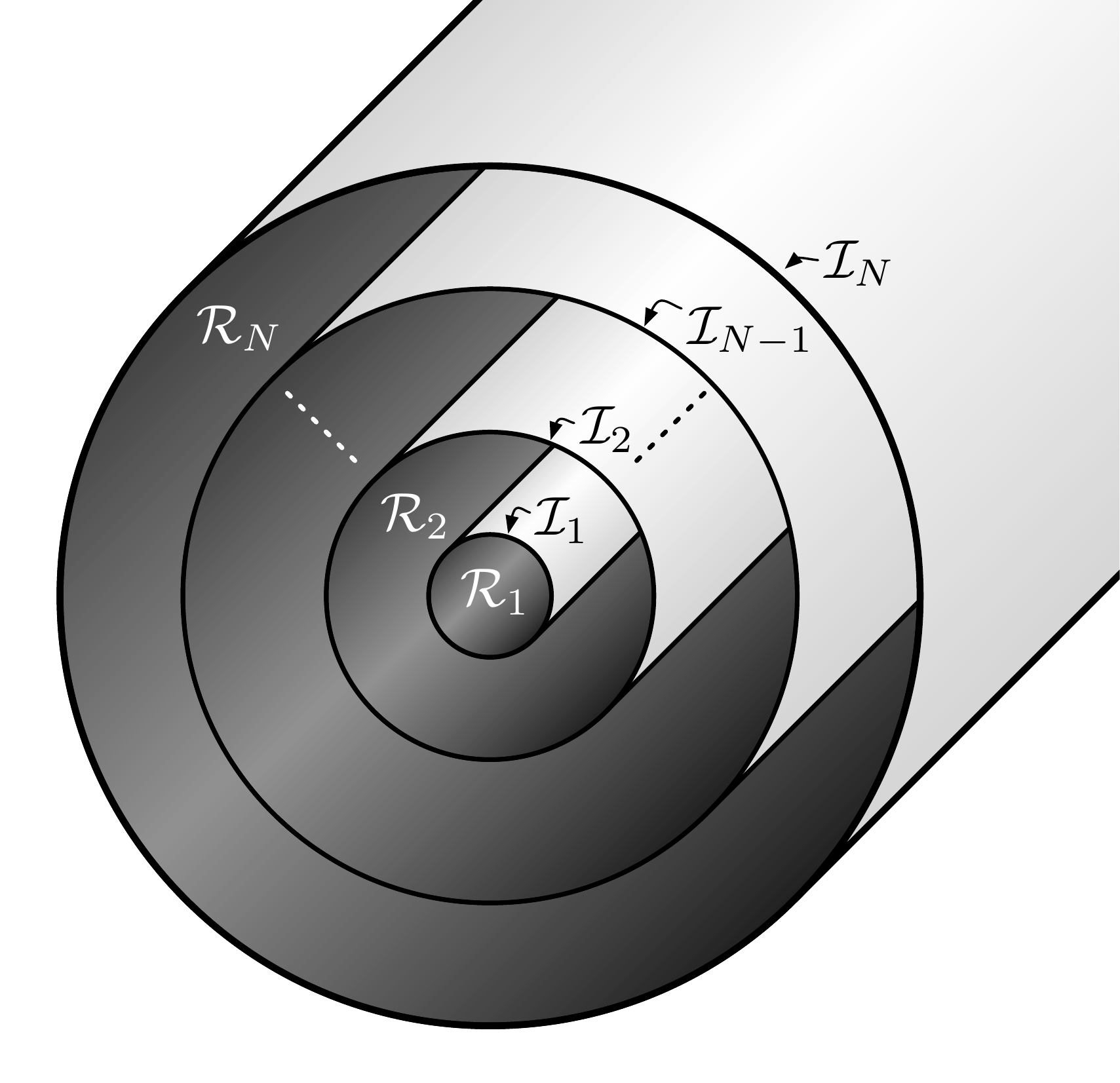}
  \caption{\label{fig:NestedSurfaces}Schematic of magnetic geometry showing ideal MHD barriers $\mathcal{I}_i$, and the relaxed plasma regions $\mathcal{R}_i$.}
\end{figure}

The model we present in this paper is an extension of the MRxMHD model introduced previously \citep{Hole:2006,Hole:2007,Hudson:2007,Dewar:2008}.  Briefly, the MRxMHD model consists of $N$ nested plasma regions $\mathcal{R}_i$ separated by ideal MHD barriers $\mathcal{I}_i$ (see Fig.~\ref{fig:NestedSurfaces}).  Each plasma region is assumed to have undergone Taylor relaxation \citep{Taylor:1986} to a minimum energy state subject to conserved fluxes and magnetic helicity.  The MRxMHD model minimizes the plasma energy
\begin{align}
  E &= \sum_i E_i = \sum_i \int_{\mathcal{R}_i} \left(\frac{1}{2} \mathbf{B}^2 + \frac{1}{\gamma-1} p \right)\, d^3\tau, \label{eq:NoFlowPlasmaEnergy}
\end{align}
where we have used units such that $\mu_0=1$, and the minimization of Eq.~\eqref{eq:NoFlowPlasmaEnergy} is subject to constraints on the plasma mass $M_i$ and the magnetic helicity $K_i$, which are given by
\begin{align}
  M_i &= \int_{\mathcal{R}_i} \rho\, d^3\tau, \label{eq:PlasmaMass}\\
  \begin{split}
  K_i &= \int_{\mathcal{R}_i} \mathbf{A}\cdot\mathbf{B}\, d^3\tau \\
  &\relphantom{=} - \Delta\psi_{p,i} \oint_{\mathcal{C}_{p,i}^{<}} \mathbf{A} \cdot d\mathbf{l} - \Delta \psi_{t,i} \oint_{\mathcal{C}_{t,i}^{>}} \mathbf{A} \cdot d\mathbf{l},
  \end{split} \label{eq:MagneticHelicity}
\end{align}
where $p$ is the plasma pressure, $\rho$ is the plasma mass density, $\mathbf{A}$ is the magnetic vector potential, and the loop integrals in Eq.~\eqref{eq:MagneticHelicity} are required for gauge invariance.  The plasma in each volume is assumed to obey the adiabatic equation of state $\sigma_i = p/\rho^\gamma$ with $\sigma_i$ constant in each region. Additionally, each plasma region $\mathcal{R}_i$ is bounded by magnetic flux surfaces and is constrained to have enclosed toroidal flux $\Delta\psi_{t,i}$ and poloidal flux $\Delta\psi_{p,i}$.  The $\mathcal{C}_{p,i}^<$ and $\mathcal{C}_{t,i}^>$ are circuits about the inner ($<$) and outer ($>$) boundaries of $\mathcal{R}_i$ in the poloidal and toroidal directions, respectively.

Minimum energy states of the MRxMHD model are stationary points of the energy functional
\begin{align}
  W &= \sum_i \left[ E_i - \nu_i \left(M_i - M_i^0\right) - \frac{1}{2} \mu_i \left(K_i - K_i^0\right)\right] , \label{eq:NoFlowEnergyFunctional}
\end{align}
where $\nu_i$ and $\mu_i$ are Lagrange multipliers respectively enforcing the plasma mass and magnetic helicity constraints, and the $M_i^0$ and $K_i^0$ are respectively the constrained values of the plasma mass and magnetic helicity.  

Setting the first variation of Eq.~\eqref{eq:NoFlowEnergyFunctional} to zero gives \citep{Hole:2006}
\begin{align}
  \nabla \times \mathbf{B} &= \mu_i \mathbf{B}, \label{eq:NoFlowVolumeCondition1}\\
  p_i &= \text{const},  \label{eq:NoFlowVolumeCondition2}\\
  0 &= \left[\left[p_i + \frac{1}{2} \mathbf{B}^2 \right]\right], \label{eq:NoFlowInterfaceCondition}
\end{align}
where Eqs.~\eqref{eq:NoFlowVolumeCondition1}--\eqref{eq:NoFlowVolumeCondition2} apply in each plasma region $\mathcal{R}_i$, Eq.~\eqref{eq:NoFlowInterfaceCondition} applies on each ideal interface $\mathcal{I}_i$, and $\jump{x} = x_{i+1} - x_{i}$ denotes the change in the quantity $x$ across the interface $\mathcal{I}_i$.

\subsection{Including the effects of plasma flow}
\label{sec:FlowMRxMHD}

In previous work we extended the MRxMHD model to include plasma flow \citep{Dennis:2014}.  That model is defined by minimizing the plasma energy
\begin{align}
  E &= \sum_i E_i = \sum_i \int_{\mathcal{R}_i}\left(\frac{1}{2}\rho \mathbf{u}^2 + \frac{1}{2} \mathbf{B}^2 + \frac{1}{\gamma - 1}p \right)\, d^3\tau, \label{eq:FlowPlasmaEnergy}
\end{align}
where $\mathbf{u}$ is the mean plasma velocity.  The minimization of the plasma energy is subject to constraints on the plasma mass and helicity given by Eqs.~\eqref{eq:PlasmaMass}--\eqref{eq:MagneticHelicity}, and additional constraints on the flow helicity $C_i$ and toroidal angular momentum $L_i$, which are given by
\begin{align}
  C_i &= \int_{\mathcal{R}_i} \mathbf{B}\cdot\mathbf{u} \, d^3 \tau, \label{eq:FlowHelicity}\\
  L_i &= \uZ \cdot \int_{\mathcal{R}_i} \rho \mathbf{r} \times \mathbf{u} \, d^3 \tau = \int_{\mathcal{R}_i} \rho R \mathbf{u} \cdot \uphi\, d^3\tau, \label{eq:AngularMomentum}
\end{align}
where the $(R, Z, \phi)$ cylindrical coordinate system is used with $\uZ$ a unit vector pointing along the axis of symmetry, and $\phi$ the toroidal angle.

As described in detail in \citet{Dennis:2014}, constraining the toroidal angular momentum $L_i$ in each plasma region requires assuming the plasma to be axisymmetric.  A more appropriate model for 3D MHD structures is obtained if instead only the total toroidal angular momentum $L = \sum_i L_i$ is constrained\footnote{A separate derivation is not needed for this case, as the appropriate Euler-Lagrange equations are readily obtained by replacing the Lagrange multipliers $\Omega_i$ with the single Lagrange multiplier $\Omega$.  Similarly, the toroidal angular momentum constraint can be completely relaxed by the replacement $\Omega_i \rightarrow 0$.}.  This only requires the assumption that the outer plasma boundary be axisymmetric.  In the case of stellarators or other situations where the plasma boundary is not axisymmetric, the toroidal angular momentum constraint must be relaxed entirely.

In our earlier work \citep{Dennis:2014} we solved this variational problem assuming the adiabatic equation of state $p = \sigma_i \rho^\gamma$, where $\sigma_i$ is constant in each plasma region.  This is appropriate if relaxation is assumed to occur fast enough that heat transport is negligible.  An alternative approach, which was taken by \citet{Finn:1983}, is to instead maximize the plasma entropy in each region, while conserving the plasma energy, mass, helicity, flow helicity and angular momentum.  This is equivalent to assuming that parallel heat transport is rapid and that the plasma has reached thermal equilibrium along each field line.  \citet{Finn:1983} prove that the Euler-Lagrange equations (Eqs.~\eqref{eq:NoFlowVolumeCondition1}--\eqref{eq:NoFlowInterfaceCondition}) obtained from this approach are identical to those obtained by instead minimizing the plasma energy while holding the plasma entropy and other constraints fixed.  The only difference between these two approaches is that for a given initial state, the final relaxed states will be different if entropy is maximized while conserving energy versus minimizing energy while conserving entropy.  In this article, we will take the approach of minimizing energy for consistency with our earlier work \citep{Hole:2006,Hole:2007,Hudson:2007,Dewar:2008,Dennis:2013a,Dennis:2014}, however identical Euler-Lagrange equations are obtained with either approach.

The two equations of state used to complete the MRxMHD model with flow, namely assuming the adiabatic equation of state $p = \sigma_i \rho^\gamma $ or conserving the plasma entropy are described in the following sections.

\subsubsection{Adiabatic equation of state}
\label{sec:MRxMHDWithFlowAdiabatic}

If the adiabatic equation of state $p = \sigma_i \rho^\gamma$ is assumed, the minimum energy states are stationary points of the energy functional
\begin{align}
  \begin{split}
    W = \sum_i \bigg[ & E_i - \nu_i (M_i - M_i^0) - \frac{1}{2} \mu_i (K_i - K_i^0) \\
                      & - \lambda_i (C_i - C_i^0) - \Omega_i (L_i - L_i^0) \bigg],
  \end{split} \label{eq:FlowEnergyFunctional}
\end{align}
where $\lambda_i$ and $\Omega_i$ are Lagrange multipliers enforcing the flow-helicity and angular momentum constraints.

We have previously shown that the minimum energy states of this model satisfy \citep{Dennis:2014}
\begin{align}
  \nabla \times \mathbf{B} &= \mu_i \mathbf{B} + \lambda_i \nabla \times \mathbf{u}, \label{eq:Flow:Adiabatic:Field}\\
  \rho \mathbf{u} &= \lambda_i \mathbf{B} + \rho \Omega_i R \uphi, \label{eq:Flow:Adiabatic:Flow}\\
  \nu_i &= \frac{1}{2} \mathbf{u}^2 + \frac{\gamma}{\gamma - 1} \frac{p}{\rho} - \Omega_i R \mathbf{u} \cdot \uphi, \label{eq:Flow:Adiabatic:Bernoulli}\\
  p &= \sigma_i \rho^\gamma, \label{eq:Flow:Adiabatic:Pressure}\\
  0 &= \jump{\frac{1}{2} \mathbf{B}^2 + p}. \label{eq:Flow:Adiabatic:InterfaceCondition}
\end{align}
In contrast to the zero-flow limit, pressure is not constant in each plasma region, but instead there are non-zero pressure gradients.  This model was discussed in detail in our earlier work \citep{Dennis:2014}.

\subsubsection{Conservation of entropy}
\label{sec:MRxMHDWithFlowEntropy}

Instead of assuming the adiabatic equation of state, an alternative is to conserve the plasma entropy
\begin{align}
  S_i &= \int_{\mathcal{R}_i}  \frac{1}{\gamma - 1} \rho \ln \left(\frac{p}{\rho^\gamma}\right)\, d^3\tau. \label{eq:IsotropicEntropy}
\end{align}
In this case, the energy functional Eq.~\eqref{eq:FlowEnergyFunctional} gains the additional term $- \sum_i T_i (S_i - S_i^0)$, where $T_i$ is a Lagrange multiplier that will be identified as the plasma temperature.

The minimum energy states of this model satisfy
\begin{align}
  \nabla \times \mathbf{B} &= \mu_i \mathbf{B} + \lambda_i \nabla \times \mathbf{u}, \label{eq:Flow:Entropy:Field}\\
  \rho \mathbf{u} &= \lambda_i \mathbf{B} + \rho \Omega_i R \uphi, \label{eq:Flow:Entropy:Flow}\\
  \nu_i &= \frac{1}{2} \mathbf{u}^2 - \frac{T_i}{\gamma - 1} \left[\ln \left(\frac{p}{\rho^\gamma}\right) - \gamma\right] - \Omega_i R \mathbf{u} \cdot \uphi, \label{eq:Flow:Entropy:Bernoulli}\\
  p &= \rho T_i, \label{eq:Flow:Entropy:Pressure}\\
  0 &= \jump{\frac{1}{2} \mathbf{B}^2 + p}, \label{eq:Flow:Entropy:InterfaceCondition}
\end{align}
where from Eq.~\eqref{eq:Flow:Entropy:Pressure} we can identify the Lagrange multiplier $T_i$ as the plasma temperature in each region (in units where the Boltzmann constant $k_B = 1$).  The model given by Eqs.~\eqref{eq:Flow:Entropy:Field}--\eqref{eq:Flow:Entropy:InterfaceCondition} is the isotropic limit of the anisotropic MRxMHD model presented in the next section.  A derivation of that model is given in Appendix~\ref{sec:MRxMHDDerivation}.

In this model the plasma has constant temperature $T_i$ in each region.  Note that in deriving Eq.~\eqref{eq:Flow:Entropy:Pressure} have not assumed that the plasma obeys an isothermal equation of state during relaxation, as the temperature $T_i$ is not known \emph{a priori}.  Instead, the final equilibrium temperatures in each region are determined by the conservation of plasma entropy in each region, and may change from their initial values.

In the zero-flow limit, the conservation of entropy approach is equivalent to assuming the adiabatic equation of state.  In this limit the two are related by $\sigma_i = \exp\left[(\gamma - 1) S_i^0 / M_i^0\right]$.  Thus in the zero-flow limit both MRxMHD flow models reduce to the zero-flow model presented in Sec.~\ref{sec:ZeroFlowMRxMHD}.

\subsection{Including the effects of pressure anisotropy}
\label{sec:MRxMHDWithAnisotropy}

We present here an extension to MRxMHD to include the effects of pressure anisotropy.  This model is an extension to our previous work that included the effects of bulk plasma flow \citep{Dennis:2014}, and includes ideas from the work of \citet{Sun:1987}.  In our model, each plasma region is assumed to have undergone a generalized type of Taylor relaxation which minimizes the plasma energy
\begin{align}
  E &= \sum_i E_i = \sum_i \int_{\mathcal{R}_i} \left(\frac{1}{2} \rho \mathbf{u}^2 + \frac{1}{2}\mathbf{B}^2 + \frac{1}{2}p_\parallel + p_\perp \right)\, d^3\tau \label{eq:FiniteVolumePlasmaEnergy}
\end{align}
subject to constraints of the plasma mass $M_i$ (Eq.~\eqref{eq:PlasmaMass}), magnetic helicity $K_i$ (Eq.~\eqref{eq:MagneticHelicity}), flow helicity $C_i$ (Eq.~\eqref{eq:FlowHelicity}), angular momentum $L_i$ (Eq.~\eqref{eq:AngularMomentum}), and the additional quantities
\begin{align}
  S_i &= \int_{\mathcal{R}_i} \frac{1}{2} \rho \ln \left(\frac{p_\parallel p_\perp^2}{\rho^5}\right)\, d^3\tau, \label{eq:AnisotropicEntropy} \\
  G_i[F] &= \int_{\mathcal{R}_i} \rho F\left(\frac{p_\perp}{\rho B}\right)\, d^3\tau, \label{eq:IntegratedDipoleMoment}
\end{align}
where $S_i$ is the anisotropic plasma entropy, and $G_i[F]$ is a conserved quantity related to the magnetic moment of the plasma gyro-motion, which is written in terms of the unspecified function $F({p_\perp}/{\rho B})$.  Additionally, $p_\parallel$ and $p_\perp$ are the parallel and perpendicular pressures, and $B = \abs{\mathbf{B}}$ is the magnitude of the magnetic field.  The plasma quantities constrained by this model are all conserved by double-adiabatic (CGL \citep{Chew:1956}) anisotropic ideal MHD, and are assumed to be robust in the presence of small amounts of resistivity and viscosity.  The anisotropic entropy (Eq.~\eqref{eq:AnisotropicEntropy}) reduces to the isotropic entropy (Eq.~\eqref{eq:IsotropicEntropy}) in the limit $p_\parallel = p_\perp$ with $\gamma = 5/3$.

The constraints $S_i$ and $G_i$ are a generalization of the parallel and perpendicular entropies defined by \citet{Sun:1987}
\begin{align}
  S_\parallel &= \int \rho \ln \left(\frac{p_\parallel B^2}{\rho^3}\right)\, d^3\tau, \\
  S_\perp &=  \int \rho \ln \left(\frac{p_\perp}{\rho B}\right)\, d^3\tau,
\end{align}
where $S_i = \frac{1}{2} S_\parallel + S_\perp$, and $G_i = S_\perp$ with the function $F(x)$ in Eq.~\eqref{eq:IntegratedDipoleMoment} given by the choice $F(x) = \ln\left(x\right)$.  Hence, our choice of constraints $S_i$ and $G_i$ include those considered by \citet{Sun:1987}, but are more general as the function $F(x)$ is unspecified.  This unspecified function can be thought of as an anisotropic equation of state, and in Sec.~\ref{sec:ContinuousLimit:Enthalpy} is shown to be related to the anisotropic plasma enthalpy.  Another valid choice for $F(x)$ is $F(x) = x$, which corresponds to constraining the quantity $\int \left(p_\perp / B \right) \, d^3\tau $.  We show in Sec.~\ref{sec:ContinuousLimit:Enthalpy} that this choice of $F(x)$ is equivalent to the two-temperature guiding-centre plasma equation of state in anisotropic ideal MHD \citep{Iacono:1990}.

The choice to constrain the quantity $G_i$ is motivated by the magnetic moment adiabatic invariant $\tilde{\mu}$, which in the CGL anisotropic MHD model\citep{Chew:1956}, is assumed to be constant along magnetic field lines
\begin{align}
  \frac{d}{dt} \tilde{\mu} &= \frac{d}{dt} \left(\frac{p_\perp}{\rho B} \right) = 0.
\end{align}
This equation of motion corresponds to the infinity of constraints
\begin{align}
  G[F] &= \int \rho F\left(\frac{p_\perp}{\rho B}\right)\, d^3\tau,
\end{align}
for all functions $F(x)$.  The model presented in this work selects one element of this class of invariants as the most conserved of this class.  Choosing the function $F(x)$ specifies this choice, and is effectively an anisotropic equation of state.

Minimum energy states of the MRxMHD model with anisotropy and flow are stationary points of the energy functional
\begin{align}
  \begin{split}
    W = \sum_i \bigg[ &E_i - \nu_i (M_i - M_i^0) - \frac{1}{2} \mu_i(K_i - K_i^0) \\
    & -  \lambda_i (C_i - C_i^0) -  \Omega_i(L_i - L_i^0) \\
    & -  T_i(S_i - S_i^0) -  \eta_i (G_i - G_i^0) \bigg],
  \end{split} \label{eq:Anisotropic:EnergyFunctional}
\end{align}
where $\eta_i$ is a Lagrange multiplier enforcing the constraint on the quantity $G_i$.

Setting the first variation of Eq.~\eqref{eq:Anisotropic:EnergyFunctional} to zero gives the plasma region conditions
\begin{align}
  \nabla \times \mathbf{B} &= \mu_i \mathbf{B} + \lambda_i \nabla \times \mathbf{u} + \nabla \times \left[\left(\frac{p_\parallel - p_\perp}{B^2}\right) \mathbf{B}\right], \label{eq:Anisotropic:Field}\\
  \rho \mathbf{u} &= \lambda_i \mathbf{B} + \rho \Omega_i R \uphi, \label{eq:Anisotropic:Flow} \\
  \begin{split}
  \nu_i &= \frac{1}{2}\mathbf{u}^2 - \frac{1}{2} T_i \left[\ln\left(\frac{p_\parallel p_\perp^2}{\rho^5}\right) - 5\right]\\
  &\relphantom{=} - \eta_i F\left(\frac{p_\perp}{\rho B}\right) - \frac{p_\parallel - p_\perp}{\rho} - \Omega_i R \mathbf{u}\cdot \uphi, 
  \end{split} \label{eq:Anisotropic:Bernoulli} \\
  p_\parallel &= \rho T_i, \label{eq:Anisotropic:ParallelPressure} \\
  p_\perp &= \rho T_i + \eta_i \frac{p_\perp}{B} F'\left(\frac{p_\perp}{\rho B}\right), \label{eq:Anisotropic:PerpendicularPressure}
\end{align}
together with the interface force-balance condition
\begin{align}
  \left[\left[\frac{1}{2} \mathbf{B}^2 + p_\perp \right]\right] &= 0. \label{eq:Anisotropic:InterfaceForceBalance}
\end{align}
A derivation of these equations is given in Appendix~\ref{sec:MRxMHDDerivation}.

Taking the isotropic limit ($\eta = 0$) gives MRxMHD with flow with conserved entropy (see Sec.~\ref{sec:MRxMHDWithFlowEntropy}) with $\gamma = 5/3$.

In Appendix~\ref{sec:ForceBalanceProof} we show that the MRxMHD minimum energy states described by Eqs.~\eqref{eq:Anisotropic:Field}--\eqref{eq:Anisotropic:PerpendicularPressure} satisfy
\begin{align}
  \begin{split}
  \rho \left(\mathbf{u} \cdot \nabla\right) \mathbf{u} = & - \nabla \cdot \overleftrightarrow{P} + \mathbf{J} \times \mathbf{B} \\
   &- \rho \Omega_i R \uphi \times \left(\nabla \times \mathbf{u}\right) + \rho \Omega_i \nabla (R \mathbf{u} \cdot \uphi),
  \end{split} \label{eq:AnisotropicLabFrameForceBalance}
\end{align}
where $\overleftrightarrow{P}$ is the pressure tensor, which is given by
\begin{align}
  \overleftrightarrow{P} &= p_\perp \overleftrightarrow{I} + \left(p_\parallel - p_\perp\right) \mathbf{B} \mathbf{B} / B^2, \label{eq:PressureTensor}
\end{align}
with $\overleftrightarrow{I}$ the identity tensor.  Equation~\eqref{eq:AnisotropicLabFrameForceBalance} does not take the form of an equation for force-balance in the laboratory reference frame unless the last two terms on the right-hand side are zero.  Instead, Eq.~\eqref{eq:AnisotropicLabFrameForceBalance} is equivalent to force-balance in a reference frame rotating about the $Z$ axis with angular frequency $\Omega_i$.  In \citet{Dennis:2014}, a similar phenomenon was discussed for isotropic MRxMHD with flow.  Indeed our Eq.~\eqref{eq:AnisotropicLabFrameForceBalance} is identical to Eq.~(16) of \citet{Dennis:2014} with $\nabla \cdot \overleftrightarrow{P}$ replacing $\nabla p$.  The consequence of this is that the minimum energy anisotropic MRxMHD states will not necessarily be time-independent in the laboratory frame, but will be time-independent in a reference frame rotating with angular frequency $\Omega_i$.

\subsubsection{Choices for the function $F(x)$}

If the choice $F(x) = \ln(x)$ is made as in \citet{Sun:1987}, then the parallel and perpendicular temperatures are constant in each plasma region
\begin{align}
  p_\parallel &= \rho T_i, \\
  p_\perp &= \rho \left(\eta_i + T_i\right).
\end{align}
The Bernoulli equation (Eq.~\eqref{eq:Anisotropic:Bernoulli}) becomes
\begin{align}
  \begin{split}
    \nu_i = &\frac{1}{2} \mathbf{u}^2 - \frac{1}{2} T_i \left[\ln \left(\frac{T_i (T_i + \eta_i)^2}{\rho^2}\right) - 5\right]\\
    &-\eta_i \left[\ln\left(\frac{T_i + \eta_i}{B}\right) - 1\right] - \Omega_i R \mathbf{u} \cdot \uphi.
  \end{split}
\end{align}

If instead the choice $F(x) = x$ is made, then the parallel temperature is constant in each plasma region, but the perpendicular temperature depends on the magnitude of the magnetic field $B$,
\begin{align}
  p_\parallel &= \rho T_i, \label{eq:Anisotropic:F=x:PParallel}\\
  p_\perp &= \rho T_i \frac{B}{B - \eta_i}. \label{eq:Anisotropic:F=x:PPerp}
\end{align}
The Bernoulli equation (Eq.~\eqref{eq:Anisotropic:Bernoulli}) becomes
\begin{align}
  \begin{split}
    \nu_i = &\frac{1}{2} \mathbf{u}^2 - \frac{1}{2} T_i \left[\ln \left(\frac{T_i^3}{\rho^2 \left(1 - \eta_i/B\right)^2} \right) - 5\right] \\
    & - \Omega_i R \mathbf{u} \cdot \uphi.
  \end{split}
\end{align}
The pressure equations given by Eqs.~\eqref{eq:Anisotropic:F=x:PParallel}--\eqref{eq:Anisotropic:F=x:PPerp} are identical to those of the guiding-centre plasma two-temperature closure relations (see Eq.~(34) of \citet{Iacono:1990}).  It is shown in the next section that the choice $F(x)=x$ corresponds identically to this model in the continuously nested flux surface limit.

\subsubsection{Summary}

We have presented a multi-region relaxation model for plasmas which includes both anisotropy and flow.  We validate our model in Sec.~\ref{sec:ContinuousLimit} by proving that it approaches anisotropic ideal MHD with flow in the limit as the number of plasma volumes $N$ becomes large, and this is independent of the choice of the function $F(x)$.  We have previously proven that MRxMHD with flow approaches ideal MHD with flow \citep{Dennis:2014}.

\section{The continuously nested flux-surface limit}
\label{sec:ContinuousLimit}

In this section we take the continuously nested flux surface limit ($N \rightarrow \infty$) of anisotropic MRxMHD and prove that it reduces to anisotropic ideal MHD. 

Taking the limit of infinitesimally small plasma regions of the energy functional Eq.~\eqref{eq:Anisotropic:EnergyFunctional} gives
\begin{align}
  \begin{split}
  W =& \int \left(\frac{1}{2}\rho \mathbf{u}^2 + \frac{1}{2}\mathbf{B}^2 + \frac{1}{2}p_\parallel + p_\perp \right)d^3\tau\\
  & - \int \nu(s)\left(dM - dM^0\right) - \int\frac{1}{2}\mu(s)\left(dK - dK^0\right) \\
  & - \int \lambda(s) \left(dC - dC^0\right) - \int \Omega(s)\left(dL - dL^0\right) \\
  & - \int T(s) \left(dS - dS^0\right) - \int \eta(s) \left(dG - dG^0\right),
  \end{split}
  \label{eq:ContinuousEnergyFunctional1}
\end{align}
where $s$ is an arbitrary flux-surface label;  $dM$, $dK$, $dC$, $dL$, $dS$ and $dG$ are respectively infinitesimal amounts of plasma mass, magnetic helicity, flow helicity, toroidal angular momentum, plasma entropy and the magnetic dipole constraint $G$ between infinitesimally separated flux surfaces; and $dM^0$, $dK^0$, $dC^0$, $dL^0$, $dS^0$, and $dG^0$ are the corresponding constraints.

In the finite-volume limit the magnetic flux constraints are enforced by restricting the class of perturbations of the vector potential $\delta\mathbf{A}$ (see Appendix~\ref{sec:MRxMHDDerivation}), and these constraints are therefore not included in the energy functional given by Eq.~\eqref{eq:ContinuousEnergyFunctional1}.  In the limit of continuously nested flux surfaces we use the same approach we used in \citet{Dennis:2014} and introduce a vector of Lagrange multipliers $\mathbf{Q} = Q_s(s, \theta, \zeta) \nabla s + Q_\theta(s) \nabla \theta + Q_\zeta(s) \nabla \zeta$ to enforce the radial, poloidal and toroidal magnetic flux constraints in an $(s, \theta, \zeta)$ coordinate system with $\theta$ an arbitrary poloidal angle coordinate and $\zeta$ an arbitrary toroidal angle coordinate.  As detailed in Sec.~III~A of \citet{Dennis:2014}, enforcing the magnetic flux constraints requires adding the following terms to the right-hand side of the energy functional Eq.~\eqref{eq:ContinuousEnergyFunctional1}
\begin{align}
  \begin{split}
  \left. W \right|_\text{flux constraints} = & - \int \left(\mathbf{Q} \cdot \mathbf{B}\right) \, d^3 \tau \\
  & + 2\pi \int \left[ Q_\theta(s) \frac{d \psi_p^0(s)}{ds} + Q_\zeta(s) \frac{d\psi_t^0(s)}{ds} \right]\, ds,
  \end{split}
\end{align}
where $\psi_p(s)$ and $\psi_t(s)$ are respectively the poloidal and toroidal magnetic fluxes enclosed by the flux surface with label $s$.

In \citet{Dennis:2014} we showed that the magnetic helicity constraint is trivially satisfied in the limit of continuously nested flux surfaces with $dK = dK^0$ following from conservation of the magnetic fluxes within every flux surface.  Therefore the magnetic helicity term $\int \frac{1}{2} \mu(s)(dK - dK^0)\,ds$ in Eq.~\eqref{eq:ContinuousEnergyFunctional1} is zero.

With these simplifications we obtain the energy functional
\begin{align}
  \begin{split}
  W =& \int \bigg[\frac{1}{2}\rho \mathbf{u}^2 + \frac{1}{2}\mathbf{B}^2 + \frac{1}{2}p_\parallel + p_\perp - \mathbf{Q}\cdot\mathbf{B} - \nu(s) \rho \\
  & \phantom{\int\bigg[} -\lambda(s) \mathbf{B}\cdot\mathbf{u} - \rho \Omega(s) R \mathbf{u}\cdot\uphi \\
  & \phantom{\int\bigg[} - \frac{1}{2}T(s) \rho \ln\left(\frac{p_\parallel p_\perp^2}{\rho^5}\right) - \eta(s) \rho F\left(\frac{p_\perp}{\rho B}\right) \bigg]\, d^3\tau\\
  & + \int \bigg[2\pi Q_\theta(s) \frac{d \psi_p^0(s)}{ds} + 2\pi Q_\zeta(s) \frac{d\psi_t^0(s)}{ds} \\
  & \phantom{+\int\bigg[} + \nu(s) \frac{dM^0(s)}{ds} + \lambda(s) \frac{dC^0(s)}{ds} + \Omega(s) \frac{dL^0(s)}{ds}\\
  & \phantom{+\int\bigg[} + T(s) \frac{dS^0(s)}{ds} + \eta(s) \frac{dG^0(s)}{ds} \bigg]\, ds.
  \end{split}
  \label{eq:ContinuousEnergyFunctional2}
\end{align}
Requiring zero variations of $W$ with respect to the Lagrange multipliers enforce the corresponding constraints.  The interesting variations are those with respect to $p_{\parallel}$, $p_\perp$ $\rho$, $\mathbf{u}$, $\mathbf{B}$, and the position of the flux surfaces $\mathbf{x}$.

Setting the variation of $W$ with respect to $p_{\parallel}$, $p_\perp$, $\rho$, $\mathbf{u}$, and $\mathbf{B}$ to zero yield respectively
\begin{align}
  p_\parallel &= \rho T(s), \\
  p_\perp &= \rho T(s) + \eta(s) \frac{p_\perp}{B} F'\left(\frac{p_\perp}{\rho B}\right), \\
  \begin{split}
  \nu(s) &= \frac{1}{2}\mathbf{u}^2 - \frac{1}{2} T(s) \left[\ln\left(\frac{p_\parallel p_\perp^2}{\rho^5}\right) - 5\right]\\
  &\relphantom{=} - \eta(s) F\left(\frac{p_\perp}{\rho B}\right) - \frac{p_\parallel - p_\perp}{\rho} - \Omega(s) R \mathbf{u}\cdot \uphi, 
  \end{split} \label{eq:Limit:Bernoulli}\\
  \rho \mathbf{u} &= \lambda(s) \mathbf{B} + \rho \Omega(s) R \uphi, \\
  \mathbf{Q} &= \mathbf{B} - \lambda(s) \mathbf{u} - \left(\frac{p_\parallel - p_\perp}{B^2}\right) \mathbf{B}.
\end{align}

Using a very similar process to our earlier work \citep{Dennis:2014}, the variation of $W$ with respect to $\delta\mathbf{x}$ can be simplified to obtain
\begin{align}
  \begin{split}
  \left. \delta W \right|_{\delta\mathbf{x}} =  \int  \delta\mathbf{x} \cdot \big[ & \rho \left(\mathbf{u}\cdot\nabla\right) \mathbf{u} - \mathbf{J}\times\mathbf{B} + \nabla \cdot \overleftrightarrow{P}\\
  & + \rho \Omega R \uphi \times \left(\nabla \times \mathbf{u}\right) - \rho \Omega \nabla (R \mathbf{u} \cdot \uphi)  \big],
  \end{split}
\end{align}
where we have used
\begin{align}
  \begin{split}
  \nabla \cdot \overleftrightarrow{P} = &\nabla p_\perp + \mathbf{B} \left(\mathbf{B} \cdot \nabla \right) \left(\frac{p_\parallel - p_\perp}{B^2}\right) \\
  &+ \frac{p_\parallel - p_\perp}{B^2} \left(\mathbf{B} \cdot \nabla \right) \mathbf{B},
  \end{split}
  \label{eq:DivergenceOfPressureTensor}
\end{align}
which follows from the definition of the pressure tensor $\overleftrightarrow{P}$ given by Eq.~\eqref{eq:PressureTensor}.

Setting the variation $\left.\delta W\right|_{\delta\mathbf{x}}$ to zero gives
\begin{align}
  \begin{split}
  \rho \left(\mathbf{u} \cdot \nabla\right) \mathbf{u} = & - \nabla \cdot \overleftrightarrow{P} + \mathbf{J} \times \mathbf{B} \\
   &- \rho \Omega(s) R \uphi \times \left(\nabla \times \mathbf{u}\right) + \rho \Omega(s) \nabla (R \mathbf{u} \cdot \uphi ),
  \end{split}
\end{align}
which is identical to Eq.~\eqref{eq:AnisotropicLabFrameForceBalance} with the replacement $\Omega_i \rightarrow \Omega(s)$, and is an equation for force-balance in a reference frame rotating with angular velocity $\Omega(s)$ about the $\mathbf{Z}$ axis.

\subsection{The relationship between $F(x)$ and plasma enthalpy}
\label{sec:ContinuousLimit:Enthalpy}

The anisotropic ideal MHD Bernoulli equation is usually written in terms of an unspecified plasma enthalpy\footnote{See for example \citet{Iacono:1990}, where the plasma enthalpy is written as $W(\rho, B, \psi)$.} $H(\rho, B, s)$
\begin{align}
  \nu(s) &= \frac{1}{2} \mathbf{u}^2 - \Omega(s) R \mathbf{u} \cdot \uphi + H(\rho, B, s).
\end{align}
To satisfy conservation of energy, the enthalpy must satisfy the integrability conditions \citep{Iacono:1990}
\begin{align}
  \left( \frac{\partial H}{\partial \rho}\right)_{B, s} &= \frac{1}{\rho} \left(\frac{\partial p_\parallel}{\partial \rho}\right)_{B, s}, \label{eq:Limit:IntegrabilityCondition1}\\
  \left( \frac{\partial H}{\partial B}\right)_{\rho, s} &= \frac{1}{\rho} \left[\left(\frac{\partial p_\parallel}{\partial B}\right)_{\rho, s} - \frac{p_\parallel - p_\perp}{B}\right]. \label{eq:Limit:IntegrabilityCondition2}
\end{align}

By comparison with the Bernoulli equation we have derived, Eq.~\eqref{eq:Limit:Bernoulli}, we can identify the plasma enthalpy to be
\begin{align}
  \begin{split}
  H(\rho, B, s) =& -\frac{1}{2} T(s) \left[\ln \left(\frac{p_\parallel p_\perp^2}{\rho^5}\right) - 5\right]\\
  & - \eta(s) F\left(\frac{p_\perp}{\rho B}\right) - \frac{p_\parallel - p_\perp}{\rho},
  \end{split}
\end{align}
which can be shown to satisfy the integrability conditions Eqs.~\eqref{eq:Limit:IntegrabilityCondition1}--\eqref{eq:Limit:IntegrabilityCondition2} for any choice of $F(x)$.

If the function $F$ is chosen to be $F(x) = x$, then similar expressions to what were obtained in the finite plasma region limit, we obtain expressions for the plasma pressures
\begin{align}
  p_\parallel &= \rho T(s),\\
  p_\perp &= \rho T(s) \frac{B}{B - \eta(s)},
\end{align}
which are identical to the equations of state for the two-temperature guiding-centre plasma model (see \citet{Iacono:1990}).

\subsection*{Summary}

We have now proven that as the number of plasma regions $N$ becomes large in the anisotropic MRxMHD with flow model that the model reduces to anisotropic ideal MHD with flow.  The minimum energy state may not be time-independent in the laboratory reference frame, but will be time-independent in a rotating reference frame depending on the symmetry assumptions made in the model (see \citet{Dennis:2014} for details).

The energy functional given by Eq.~\eqref{eq:ContinuousEnergyFunctional2} also represents the first variational principle for anisotropic plasma equilibria with general flow fields.  This variational principle can be considered to be a generalization of that for isotropic plasma equilibria with flow described by \citet{Hameiri:1998}.

In the next section we provide a simple example calculation using our anisotropic MRxMHD model.

\section{Example application}
\label{sec:Example}

In this section, we apply our anisotropic MRxMHD model to an RFP-like plasma in the zero-flow limit.  Our example calculation is motivated by the experimental results of \citet{Sasaki:1997}, who observed ion temperature anisotropy in the EXTRAP-T2 reversed-field pinch.  In their work, \citeauthor{Sasaki:1997} measured the parallel ion temperature to be 1-3 times larger than the perpendicular temperature.  Anisotropic plasma pressures have also been observed on MST during reconnection events \citep{Magee:2011}, however on that experiment, the perpendicular temperature was observed to be greater.  In this example, we focus on the results of the EXTRAP-T2 experiment.


We model EXTRAP-T2 experiment of \citeauthor{Sasaki:1997} with single-volume anisotropic MRxMHD with zero plasma flow.  Additionally we choose $F(x) = \ln(x)$ in Eq.~\eqref{eq:IntegratedDipoleMoment} as this yields a constant ratio of parallel to perpendicular temperature, which accords with the analysis of \citeauthor{Sasaki:1997}.  In this limit, the anisotropic MRxMHD equations (Eqs.~\eqref{eq:Anisotropic:Field}--\eqref{eq:Anisotropic:PerpendicularPressure}) in SI units are
\begin{align}
  \nabla \times \mathbf{B} =& \mu \mathbf{B} - k_B \eta \nabla \times \left(\frac{\mu_0 \rho}{B^2} \mathbf{B}\right), \\
  \rho =& \rho_0 \left(\frac{B}{B_0}\right)^{-\eta/T} , \label{eq:Example:Density}\\
  p_\parallel =& \rho k_B T, \\
  p_\perp =& \rho k_B \left(T + \eta \right),
\end{align}
where $\rho_0$ is a constant reference density, $B_0$ is a constant reference magnetic field, and $k_B$ is Boltzmann's constant.

\begin{figure}
  \includegraphics[width=8cm]{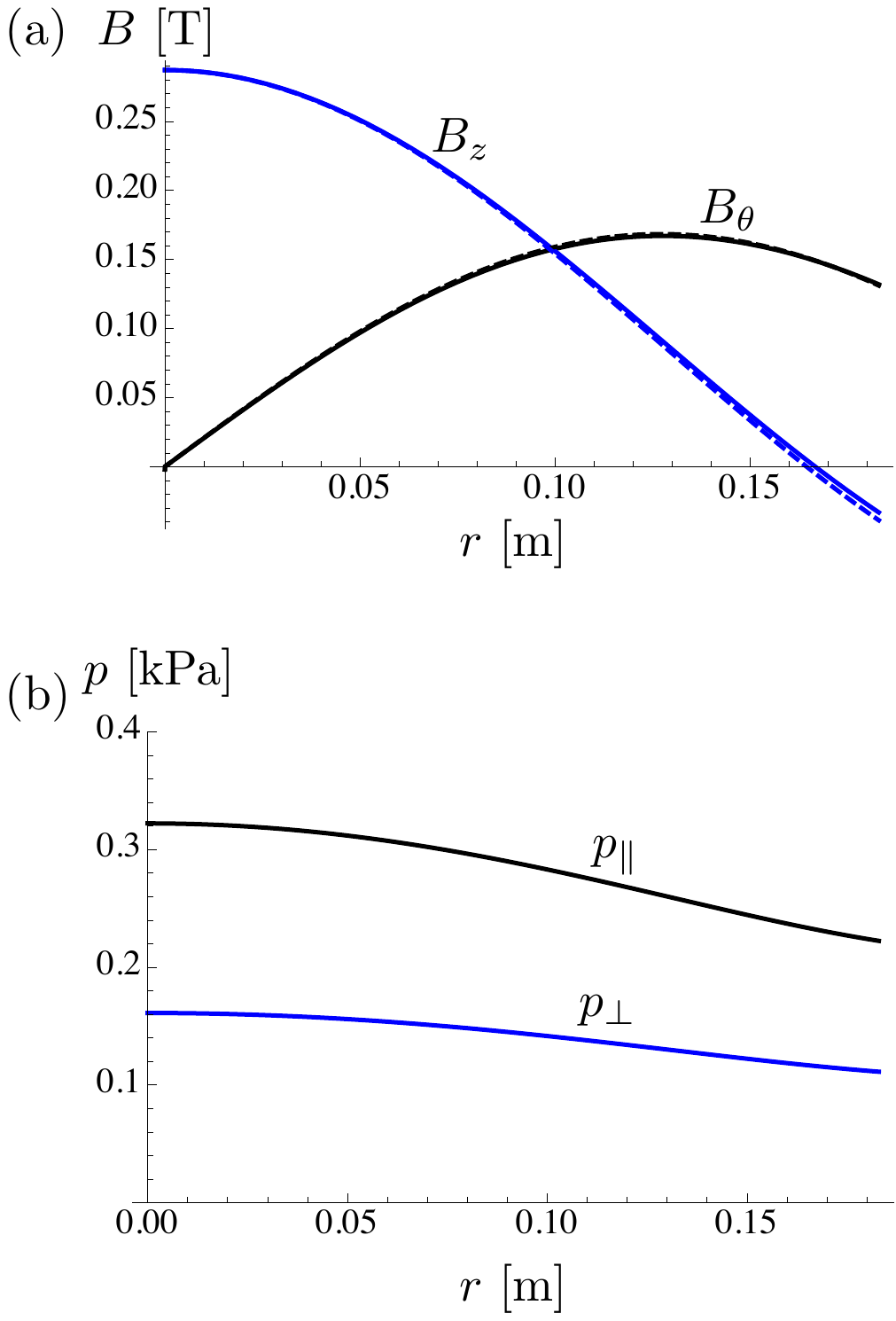}
  \caption{\label{fig:Example} Example anisotropic MRxMHD solution for an RFP in cylindrical geometry with a single plasma volume.  Panels (a) and (b), respectively, show the magnetic field components and plasma pressure versus radial position.  The dashed lines in panel (a) indicate the magnetic field profile expected if the pressure were assumed to be isotropic ($\eta = 0$).}
\end{figure}

Figure~\ref{fig:Example} illustrates the results of this model.  The equilibrium is described by $\mu = 14.4\, \text{m}^{-1}$, $T = 250\, \text{eV}$, $\eta = -170\, \text{eV}$, $\rho_0 = 8.9\times 10^{19}\,\text{m}^{-3}$, with $B_0 = 1\,\text{T}$.  These values have been chosen to ensure that the model agrees with the average experimental parameters observed during $t\approx 7\text{--}9\,\text{ms}$ in Figure~2 of \citet{Sasaki:1997}, namely major radius $R = 1.24\,\text{m}$, minor radius $a = 0.183\,\text{m}$, plasma current $I_p \approx 120\,\text{kA}$, reversal parameter $F \approx -0.4$, on-axis electron number density $\rho_e \approx 1.9 \times 10^{19}\,\text{m}^{-3}$, parallel temperature $T_\parallel \approx 250\,\text{eV}$, perpendicular temperature $T_\perp \approx 80\,\text{eV}$.  

A significant difference from the isotropic zero-flow limit presented in Sec.~\ref{sec:ZeroFlowMRxMHD} is that although the parallel and perpendicular temperatures are constant in each region, the pressures are not due to the variation of the plasma density with magnetic field strength $B$ given by Eq.~\eqref{eq:Example:Density}.  In the isotropic limit, the plasma density becomes independent of the magnetic field strength, and the pressure becomes constant in each plasma region, in agreement with Eq.~\eqref{eq:NoFlowVolumeCondition2}.

\section{Conclusion}
\label{sec:Conclusion}

We have formulated an energy principle for equilibria that comprise multiple Taylor-relaxed plasma regions including the effects of plasma anisotropy \& flow.  This model is an extension of our earlier work that considered the isotropic finite-flow limit \citep{Dennis:2014}, and the work of \citet{Sun:1987} who considered a special case of the single relaxed-region anisotropic zero-flow limit.  We have demonstrated our model reduces to anisotropic ideal MHD with flow in the limit of an infinite number of plasma regions.  This limit demonstrates the validity of our anisotropic MRxMHD model, and is to our knowledge the first variational principle for anisotropic plasma equilibria with general flow fields.  The numerical solution to the anisotropic MRxMHD model with flow presented in this work will be the subject of future work as an extension to the Stepped Pressure Equilibrium Code (SPEC) \citep{Hudson:2012}.  Implementation of the anisotropic MRxMHD model into SPEC will enable detailed comparisons between the predictions of our model in the case of fully 3D plasmas with multiple relaxed-regions and high-performance anisotropic tokamak discharges.

The authors gratefully acknowledge support of the U.S. Department of Energy and the Australian Research Council, through Grants No.\ DP0452728, No.\ FT0991899, and No.\ DP110102881.  

\appendix

\section{Derivation of the MRxMHD equations}
\label{sec:MRxMHDDerivation}

In this appendix we derive the Euler-Lagrange equations for the plasma, Eqs.~\eqref{eq:Anisotropic:Field}--\eqref{eq:Anisotropic:InterfaceForceBalance}.  The anisotropic plasma equations for a single volume have been obtained previously by \citet{Sun:1987} in the zero-flow limit and taking the function $F(x)= \ln(x)$ in the magnetic dipole constraint $G$ (see Eq.~\eqref{eq:IntegratedDipoleMoment}).  Here we extend that work by considering multiple nested volumes,  arbitrary functions $F(x)$, and including the effects of plasma flow.  Our derivation is a generalization of our earlier work \citep{Dennis:2014} to include anisotropy.

Equilibria of the anisotropic MRxMHD model are stationary points of the energy functional Eq.~\eqref{eq:Anisotropic:EnergyFunctional},
\begin{align}
  \begin{split}
    W = \sum_i \bigg[ &E_i - \nu_i (M_i - M_i^0) - \frac{1}{2} \mu_i(K_i - K_i^0) \\
    & -  \lambda_i (C_i - C_i^0) -  \Omega_i(L_i - L_i^0) \\
    & -  T_i(S_i - S_i^0) -  \eta_i (G_i - G_i^0) \bigg],
  \end{split} \label{eq:Appendix:EnergyFunctional}
\end{align}
where $\nu_i$, $\mu_i$, $\lambda_i$, $\Omega_i$, $T_i$, and $\eta_i$ are Lagrange multipliers and $E_i$, $M_i$, $K_i$, $C_i$, $L_i$, $S_i$, and $G_i$ are defined in Sec.~\ref{sec:MRxMHDModel}.

Instead of introducing Lagrange multipliers to enforce the toroidal and poloidal flux constraints as in Sec.~\ref{sec:ContinuousLimit}, we use the approach of \citet{Spies:2001} who showed that the flux constraints are equivalent to the following relationship at the interfaces
\begin{align}
  \mathbf{n} \times \delta \mathbf{A} &= - \left(\mathbf{n} \cdot \delta\mathbf{x}\right) \mathbf{B}, \label{eq:DeltaAInterfaceConstraint}
\end{align}
where $\mathbf{n}$ is a unit normal vector perpendicular to the interface boundary, $\delta\mathbf{A}$ is the variation of the vector potential, and $\delta \mathbf{x}$ is the perturbation to the interface positions.

Setting the variations of $W$ with respect to $\mathbf{u}$, $\rho$, $p_\parallel$, and $p_\perp$ to zero yield respectively
\begin{align}
  \rho \mathbf{u} &= \lambda_i \mathbf{B} + \rho \Omega_i R \uphi, \label{eq:Appendix:Flow} \\
  \begin{split}
  \nu_i &= \frac{1}{2}\mathbf{u}^2 - \frac{1}{2} T_i \left[\ln\left(\frac{p_\parallel p_\perp^2}{\rho^5}\right) - 5\right]\\
  &\relphantom{=} - \eta_i \left[ F\left(\frac{p_\perp}{\rho B}\right) - \frac{p_\perp}{\rho B} F'\left(\frac{p_\perp}{\rho B}\right) \right]  - \Omega_i R \mathbf{u}\cdot \uphi, 
  \end{split} \label{eq:Appendix:Bernoulli} \\
  p_\parallel &= \rho T_i, \label{eq:Appendix:ParallelPressure} \\
  p_\perp &= \rho T_i + \eta_i \frac{p_\perp}{B} F'\left(\frac{p_\perp}{\rho B}\right), \label{eq:Appendix:PerpendicularPressure}
\end{align}
which are equivalent to Eqs.~\eqref{eq:Anisotropic:Flow}--\eqref{eq:Anisotropic:PerpendicularPressure}.

\begin{widetext}
The variation of $W$ with respect to $\mathbf{A}$ is
\begin{align}
  \begin{split}
  \left. \delta W \right|_{\delta \mathbf{A}} = & \sum_i \int_{\mathcal{R}_i} \delta \mathbf{A} \cdot \left\{ \nabla \times \mathbf{B} - \lambda_i \nabla \times \mathbf{u} - \mu_i \mathbf{B} + \eta_i \nabla \times \left[ \frac{p_\perp}{B^3} F'\left(\frac{p_\perp}{\rho B}\right)\right] \right\} \\
  &- \sum_i \oint_{\delta \mathcal{R}_i} \left(\mathbf{n}\cdot\delta\mathbf{x}\right) \left[\mathbf{B}^2 - \frac{1}{2} \mu_i \mathbf{A} \cdot \mathbf{B}  - \lambda_i \mathbf{u} \cdot \mathbf{B} + \eta_i \frac{p_\perp}{B} F'\left(\frac{p_\perp}{\rho B}\right) \right]
  \end{split} \label{eq:Appendix:deltaWdeltaA}
\end{align}
where $\partial \mathcal{R}_i = \mathcal{I}_{i-1} \cup \mathcal{I}_i$ is the boundary of the plasma volume $\mathcal{R}_i$, and $\mathcal{I}_i$ is the plasma interface separating plasma volumes $\mathcal{R}_{i-1}$ and $\mathcal{R}_i$ (see Figure~\ref{fig:NestedSurfaces}).  The magnetic flux boundary condition, Eq.~\eqref{eq:DeltaAInterfaceConstraint}, has also been used in Eq.~\eqref{eq:Appendix:deltaWdeltaA} to write the variation of the vector potential $\delta \mathbf{A}$ on the interfaces in terms of the variation to the plasma interfaces $\delta \mathbf{x}$.

Requiring $\left. \delta W \right|_{\delta \mathbf{A}}$ to be zero for all choices of $\delta \mathbf{A}$ yields
\begin{align}
  \nabla \times \mathbf{B} &= \mu_i \mathbf{B} + \lambda_i \nabla \times \mathbf{u} - \eta_i \nabla \times \left[\frac{p_\perp}{B^3} F'\left(\frac{p_\perp}{\rho B}\right)\right], 
\end{align}
which is identical to Eq.~\eqref{eq:Anisotropic:Field} upon using the identity
\begin{align}
  \frac{p_\parallel - p_\perp}{B^2} &= - \eta_i \frac{p_\perp}{B^3} F'\left(\frac{p_\perp}{\rho B}\right), \label{eq:Appendix:DeltaPIdentity}
\end{align}
which follows from Eqs.~\eqref{eq:Appendix:ParallelPressure}--\eqref{eq:Appendix:PerpendicularPressure}.

  The interface condition can now be obtained by considering the variation of $W$ with respect to the interface positions
  \begin{align}
    \begin{split}
      \left. \delta W \right|_{\delta \mathbf{x}} =& \phantom{-}\sum_i \oint_{\partial \mathcal{R}_i} \left(\mathbf{n} \cdot \delta \mathbf{x}\right) \bigg[\frac{1}{2} \rho \mathbf{u}^2 + \frac{1}{2} \mathbf{B}^2 + \frac{1}{2} p_\parallel + p_\perp - \nu_i \rho - \lambda_i \mathbf{B} \cdot \mathbf{u} - \rho \Omega_i R \mathbf{u} \cdot \uphi - \frac{1}{2} \mu_i \mathbf{A}\cdot\mathbf{B} \\
      & \phantom{-\sum_i \oint_{\partial \mathcal{R}_i} \left(\mathbf{n} \cdot \delta \mathbf{x}\right) \bigg[} - \frac{1}{2}T_i \rho \ln \left( \frac{p_\parallel p_\perp^2}{\rho^5}\right) - \eta_i \rho F \left( \frac{p_\perp}{\rho B}\right) \bigg] \\
      & - \sum_i \oint_{\partial \mathcal{R}_i} \left(\mathbf{n} \cdot \delta \mathbf{x} \right) \left[\mathbf{B}^2 - \frac{1}{2} \mu_i \mathbf{A} \cdot \mathbf{B}  - \lambda_i \mathbf{u} \cdot \mathbf{B} + \eta_i \frac{p_\perp}{B} F'\left(\frac{p_\perp}{\rho B}\right) \right],
    \end{split} \label{eq:Appendix:deltaWdeltaX}
  \end{align}
  where the remaining term of Eq.~\eqref{eq:Appendix:deltaWdeltaA} has been included.
\end{widetext}

Equation~\eqref{eq:Appendix:deltaWdeltaX} simplifies to
\begin{align}
  \left. \delta W \right|_{\delta \mathbf{x}} &= \sum_i \oint_{\mathcal{I}_i} \left(\mathbf{n} \cdot \delta \mathbf{x}\right) \jump{\frac{1}{2}\mathbf{B}^2 + p_\perp},
\end{align}
where $\jump{x} = x_{i+1} - x_{i}$ is the jump in $x$ across the plasma interface $\mathcal{I}_i$.  Requiring this variation to be zero gives the interface condition Eq.~\eqref{eq:Anisotropic:InterfaceForceBalance},
\begin{align}
  \jump{\frac{1}{2}\mathbf{B}^2 + p_\perp} &= 0.
\end{align}

\section{Proof that MRxMHD solutions satisfy anisotropic force-balance}
\label{sec:ForceBalanceProof}

In this appendix we show that the minimum energy MRxMHD states described by the Euler-Lagrange equations, Eqs.~\eqref{eq:Anisotropic:Field}--\eqref{eq:Anisotropic:InterfaceForceBalance}, satisfy the anisotropic rotating-frame force-balance condition Eq.~\eqref{eq:AnisotropicLabFrameForceBalance}.

The magnetic field in each plasma region obeys Eq.~\eqref{eq:Anisotropic:Field}, which is
\begin{align}
  \nabla \times \mathbf{B} &= \mu_i \mathbf{B} + \lambda_i \nabla \times \mathbf{u} + \nabla \times \left[ \left(\frac{p_\parallel - p_\perp}{B^2}\right) \mathbf{B} \right].
\end{align}
Taking the cross-product of this with $\mathbf{B}$ yields
\begin{align}
  \begin{split}
  \mathbf{J} \times \mathbf{B} =& -\lambda_i \mathbf{B} \times \left(\nabla \times \mathbf{u}\right) \\
  &- \mathbf{B} \times \left\{\nabla \times \left[\left(\frac{p_\parallel - p_\perp}{B^2}\right) \mathbf{B} \right] \right\}. \label{eq:FBAppendix:JxB1}
  \end{split}
\end{align}

The first term on the right-hand side of Eq.~\eqref{eq:FBAppendix:JxB1} can be simplified using Eq.~\eqref{eq:Anisotropic:Flow} to give
\begin{align}
  \begin{split}
  -\lambda_i \mathbf{B} \times \left( \nabla \times \mathbf{u} \right) =& \rho \Omega_i R \uphi \times \left( \nabla \times \mathbf{u}\right) \\
  & - \frac{1}{2} \rho \nabla \mathbf{u}^2 + \rho \left(\mathbf{u} \cdot \nabla \right) \mathbf{u}.
  \end{split} \label{eq:FBAppendix:BCurlU}
\end{align}
Substitution back into Eq.~\eqref{eq:FBAppendix:JxB1} gives
\begin{align}
  \begin{split}
    \rho \left(\mathbf{u} \cdot \nabla \right) \mathbf{u} = & \mathbf{J} \times \mathbf{B} - \rho \Omega_i R \uphi \times \left(\nabla \times \mathbf{u}\right) + \frac{1}{2}\rho \nabla \mathbf{u}^2 \\
    &+ \mathbf{B} \times \left\{\nabla \times \left[\left(\frac{p_\parallel - p_\perp}{B^2}\right) \mathbf{B} \right] \right\}.
  \end{split}
  \label{eq:FBAppendix:JxB2}
\end{align}

Next we need to use the Bernoulli equation, Eq.~\eqref{eq:Anisotropic:Bernoulli} to write the $\rho \nabla \mathbf{u}^2$ term in Eq.~\eqref{eq:FBAppendix:JxB2} as an expression involving the divergence of the pressure tensor.  Using Eq.~\eqref{eq:Anisotropic:ParallelPressure}, the Bernoulli equation can be written as
\begin{align}
  \begin{split}
    \nu_i = & \frac{1}{2} \mathbf{u}^2 - \frac{1}{2} T_i \left[\ln\left(\frac{T_i p_\perp^2}{\rho^4} - 5\right)\right] \\
    &- \eta_i F\left(\frac{p_\perp}{\rho B}\right) - T_i + \frac{p_\perp}{\rho} - \Omega_i R \mathbf{u} \cdot \uphi.
  \end{split}
  \label{eq:FBAppendix:Bernoulli}
\end{align}
We take the gradient of the Bernoulli equation to obtain an expression involving $\rho \nabla \mathbf{u}^2$,
\begin{align}
  \begin{split}
    0 = & \frac{1}{2} \nabla \mathbf{u}^2 - \frac{1}{2} T_i \left( 2 \frac{\nabla p_\perp}{p_\perp} - 4 \frac{\nabla \rho}{\rho}\right) \\
    &- \eta_i F'\left(\frac{p_\perp}{\rho B}\right) \left(\frac{p_\perp}{\rho B}\right) \left(\frac{\nabla p_\perp}{p_\perp} - \frac{\nabla \rho}{\rho} - \frac{\nabla B}{B}\right)\\
    & + \frac{\nabla p_\perp}{\rho} - p_\perp \frac{\nabla \rho}{\rho^2} - \Omega_i \nabla (R \mathbf{u} \cdot \uphi).
  \end{split}
\end{align}
Using Eq.~\eqref{eq:Appendix:DeltaPIdentity} to rewrite $F'$ in terms of physical quantities gives
\begin{align}
  \begin{split}
    \frac{1}{2} \nabla \mathbf{u}^2 =&  T_i \left( \frac{\nabla p_\perp}{p_\perp} - 2 \frac{\nabla \rho}{\rho} \right) + \frac{p_\perp}{\rho^2} \nabla \rho - \frac{\nabla p_\perp}{\rho} \\
    & - \frac{p_\parallel - p_\perp}{\rho} \left( \frac{\nabla p_\perp}{p_\perp} - \frac{\nabla \rho}{\rho} - \frac{\nabla B}{B} \right) \\
    & + \Omega_i \nabla (R \mathbf{u} \cdot \uphi),
  \end{split}
\end{align}
which can be simplified to
\begin{align}
  \frac{1}{2} \rho \nabla \mathbf{u}^2  &= -\nabla p_\parallel + \left(\frac{p_\parallel - p_\perp}{ B^2} \right) \frac{1}{2} \nabla \mathbf{B}^2 + \rho \Omega_i \nabla (R \mathbf{u} \cdot \uphi). \label{eq:FBAppendix:GradientUSquared}
\end{align}

Equation~\eqref{eq:FBAppendix:GradientUSquared} can now be used to eliminate the $\rho \nabla \mathbf{u}^2$ term from Eq.~\eqref{eq:FBAppendix:JxB2} to give
\begin{align}
  \begin{split}
    \rho \left(\mathbf{u} \cdot \nabla\right) \mathbf{u} =& \mathbf{J} \times \mathbf{B} - \rho \Omega_i R \uphi \times \left( \nabla \times \mathbf{u}\right) + \rho \Omega_i \nabla (R \mathbf{u} \cdot \uphi) \\
    & - \nabla p_\parallel + \left(\frac{p_\parallel - p_\perp}{B^2}\right) \frac{1}{2} \nabla \mathbf{B}^2 \\
    & + \mathbf{B} \times \left\{\nabla \times \left[\left(\frac{p_\parallel - p_\perp}{B^2}\right) \mathbf{B} \right] \right\}.
  \end{split} \label{eq:FBAppendix:JxB3}
\end{align}
This is almost in the desired form of Eq.~\eqref{eq:AnisotropicLabFrameForceBalance}, all that remains to be shown is that the terms on the second and third lines of Eq.~\eqref{eq:FBAppendix:JxB3} are equal to $-\nabla \cdot \overleftrightarrow{P}$.

\begin{widetext}
The last term of Eq.~\eqref{eq:FBAppendix:JxB3} can be simplified to give
\begin{align}
  \mathbf{B} \times \left\{\nabla \times \left[\left(\frac{p_\parallel - p_\perp}{B^2}\right) \mathbf{B} \right] \right\} &=
  \left(\frac{p_\parallel - p_\perp}{B^2}\right) \mathbf{B} \times \left(\nabla \times \mathbf{B}\right) + \mathbf{B} \times \left[\nabla \left(\frac{p_\parallel - p_\perp}{B^2}\right) \times \mathbf{B}\right], \notag
\\
  &= \left(\frac{p_\parallel - p_\perp}{B^2}\right) \mathbf{B} \times \left(\nabla \times \mathbf{B}\right)  + \mathbf{B}^2 \nabla \left(\frac{p_\parallel - p_\perp}{B^2}\right) - \mathbf{B} \left(\mathbf{B}\cdot\nabla\right) \left(\frac{p_\parallel - p_\perp}{B^2}\right), \notag \\
  &= \left(\frac{p_\parallel - p_\perp}{B^2}\right) \left[\mathbf{B} \times \left(\nabla \times \mathbf{B}\right) - \nabla \mathbf{B}^2 \right]  +  \nabla \left(p_\parallel - p_\perp\right)  - \mathbf{B} \left(\mathbf{B}\cdot\nabla\right) \left(\frac{p_\parallel - p_\perp}{B^2}\right), \notag \\
  &= -\left(\frac{p_\parallel - p_\perp}{B^2}\right) \left[ \left(\mathbf{B} \cdot \nabla\right) \mathbf{B} +  \frac{1}{2} \nabla \mathbf{B}^2\right]  +  \nabla \left(p_\parallel - p_\perp\right)  - \mathbf{B} \left(\mathbf{B}\cdot\nabla\right) \left(\frac{p_\parallel - p_\perp}{B^2}\right). \label{eq:FBAppendix:BCurlZ}
\end{align}
Using this to replace the last term in Eq.~\eqref{eq:FBAppendix:JxB3} gives
\begin{align}
    \rho \left(\mathbf{u} \cdot \nabla\right) \mathbf{u} &= \mathbf{J} \times \mathbf{B} - \rho \Omega_i R \uphi \times \left( \nabla \times \mathbf{u}\right) + \rho \Omega_i \nabla (R \mathbf{u} \cdot \uphi) - \nabla p_\perp - \left(\frac{p_\parallel - p_\perp}{B^2}\right) \left(\mathbf{B}\cdot\nabla\right) \mathbf{B} - \mathbf{B} \left(\mathbf{B}\cdot\nabla\right) \left(\frac{p_\parallel - p_\perp}{B^2}\right),
\end{align}
where the last three terms are equal to $-\nabla \cdot \overleftrightarrow{P}$ (see Eq.~\eqref{eq:DivergenceOfPressureTensor}).  

We have now shown that the minimum energy MRxMHD states satisfy the anisotropic rotating-frame force-balance condition
\begin{align}
    \rho \left(\mathbf{u} \cdot \nabla\right) \mathbf{u} &= \mathbf{J} \times \mathbf{B}  - \nabla \cdot \overleftrightarrow{P} - \rho \Omega_i R \uphi \times \left( \nabla \times \mathbf{u}\right) + \rho \Omega_i \nabla (R \mathbf{u} \cdot \uphi).
\end{align}
As shown in \citet{Dennis:2014}, the last two terms of this force-balance condition mean that the plasma may not be time-independent in the laboratory frame, but will be time-independent in a reference frame rotating about the $\uZ$ axis with angular velocity $\Omega_i$.

\end{widetext}

\bibliography{MRxMHD-Anisotropy.bib}

\end{document}